# Synthesis and Structure of the Monolayer Hydrate $K_{0.3}CoO_2 \cdot 0.4H_2O$

S. Park, Y. Lee, W. Si, and T. Vogt[*]

Physics Department, Brookhaven National Laboratory, Upton, NY 11973-5000

The monolayer hydrate (MLH) $K_{0.3}CoO_2 \cdot 0.4H_2O$ was synthesized from $K_{0.6}CoO_2$ by extracting $K^+$ cations using $K_2S_2O_8$ as an oxidant and the subsequent intercalation of water between the layers of edge-sharing $CoO_6$ octahedra. A hexagonal structure (space group $P6_3/mmc$) with lattice parameters a = 2.8262(1) Å, c = 13.8269(6) Å similar to the MLH $Na_{0.36}CoO_2 \cdot 0.7H_2O$ was established using high-resolution synchrotron X-ray powder diffraction data. The $K/H_2O$ layer in the K-MLH is disordered, which is in contrast to the Na-MLH. At low temperatures metallic and paramagnetic behavior was found.

E-mail address: tvogt@bnl.gov

# 1. Introduction

Layered hydrates of metal oxides and chalcogenides with the general formula $A^{n+}_{x/n}(H_2O)_y[MX_2]^{x-}$ (A = alkali and alkaline earth metal, M = transition metal, X = O,S) have low dimensional structures composed of covalent 1- and 2-dimensional $[MX_2]^{x-}$ inorganic polymers separated by alkaline or alkaline earth cations. The high mobility of these cations at room temperature allows topotactic exchange with polar solvents, notably water to take place. Superconductivity near 5K was discovered in the Nb and Ta members of $A^{n+}_{x/n}(H_2O)_y[MS_2]^{x-}$ by Schöllhorn et al [1]. The superconductivity depends on both the A cation and the intercalated solvent. Both bilayer (BLH) and monolayer hydrates (MLH) were shown to be superconducting. The relationship of the interlayer spacing $d$ of layer-hydrate sulfides $A^{n+}_{x/n}(H_2O)_y[MS_2]^{x-}$ (A = alkali metal, M = Ti, Nb, Ta) on the charge vs. ionic radius ratio of exchangeable cations $A^{n+}$ was discussed in detail by Lerf and co-worker [2]. In contrast the structural chemistry and physical properties of layered metal oxyhydrates $A^{n+}_{x/n}(H_2O)_y[MO_2]^{x-}$ have remained largely unknown. However, the observation of superconductivity in the bilayer hydrate $Na_{0.3}CoO_2 \cdot 1.4H_2O$ [3,4] has led to an invigorated exploration of this class of materials. The monolayer hydrates $Na_{0.36}CoO_2 \cdot 0.7H_2O$ [5], $Na_{0.3}RhO_2 \cdot 0.6H_2O$ [6], and $Na_{0.22}RuO_2 \cdot 0.45H_2O$ [7] have been identified. These crystal structures are based on the stacking of two-dimensional $MO_2$ (M = Co, Rh, Ru) layers separated by a single layer containing $Na^+$ cations and $H_2O$ molecules (Fig. 1). The symmetry of $Na_{0.36}CoO_2 \cdot 0.7H_2O$ is hexagonal ($P6_3/mmc$ space group) with lattice parameters a = 2.8344(7), c = 13.842(5) Å. Both $Na_{0.3}RhO_2 \cdot 0.6H_2O$ and $Na_{0.22}RuO_2 \cdot 0.45H_2O$ have rhombohedral symmetry (R-3mH space group) with lattice parameters a = 3.0542(1), c =20.8560(9) and a = 2.930(2), c = 21.913(5) Å, respectively. In marked contrast to the above mentioned $A^{n+}_{x/n}(H_2O)_y[MS_2]^{x-}$ systems superconductivity above 2K was not observed in any MLH structures nor are there any members known with $A^{2+}$ cations. We have embarked on the exploration of the structural chemistry of the $A^{n+}_{x/n}(H_2O)_y[MO_2]^{x-}$ family of materials and present here the structure, electrical and magnetic properties of the K-analogue of the MLH cobalt oxide [8]. This layered $K_{0.3}CoO_2 \cdot 0.4H_2O$ can be synthesized from $K_{0.6}CoO_2$ [9,10,11] by extracting $K^+$ cations and intercalating water molecules using $K_2S_2O_8$ [eq.

(1)]. The extraction of K$^+$ cations leads to the oxidation of the metal in persulfate-based oxidants.

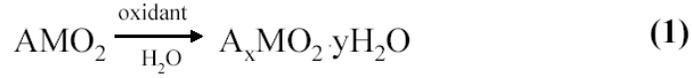

$$AMO_2 \xrightarrow[H_2O]{oxidant} A_xMO_2 \cdot yH_2O \qquad (1)$$

The synthesis, structure and some physical properties of $K_{0.3}CoO_2 \cdot 0.4H_2O$ will be discussed in the following.

## 2. Experimental

$K_{0.3}CoO_2 \cdot 0.4H_2O$ powder was prepared from the parent compound $K_{0.6}CoO_2$ using $K_2S_2O_8$ in an aqueous solution (pH~10.5) or alternatively water and stirred for 1 and 5 days respectively in a Pyrex$^{TM}$ bottle. The precursor, $K_{0.6}CoO_2$, was obtained by heating the mixture of $K_2O_2$ (12 mol % excess; Alfa 96.5%) and $Co_3O_4$ (Alfa 99.7%) at 800°C for 1 day in air. $K_{0.6}CoO_2$ with excess $K_2S_2O_8$ (molar ratio = 1:2) was placed in 20 mL DI water. After adding 4 drops of 1 N $NH_4OH$ (pH ~10.5) in a beaker covered with a Parafilm$^{TM}$ the solution was stirred for one day, filtered, and dried in air. Phase identification was established using a MiniFlex$^{TM}$ (Rigaku) diffractometer (Cu $K\alpha$ radiation) and the structrure was refined using the high-resolution synchrotron X-ray powder diffraction data measured at beam line X7A at the National Synchrotron Light Source at Brookhaven National Laboratory. The $K_{0.3}CoO_2 \cdot 0.4H_2O$ was contained in a humidity chamber prior to the experiment (constant exposure to water vapor pressure at room temperature). A powder sample of $K_{0.3}CoO_2 \cdot 0.4H_2O$ was loaded into a 0.5mm glass capillary. The capillary was subsequently sealed and mounted on the 2nd axis of the diffractometer. A monochromatic beam of 0.6981 Å was selected via a channel-cut Ge(111) monochromator. A gas-propotional position-sensitive detector (PSD), gated at the Kr-escape peak, was employed for high-resolution ($\Delta d/d \approx 10^{-3}$) powder diffraction data measurements [12]. The PSD was stepped in 0.25° intervals between 5° and 60°. The compositions of the systems K-Co and their water contents were determined by inductively coupled plasma atomic-emission spectrometry (ICP-AES) and thermogravimetric analysis (TGA), respectively. Magnetization measurements were carried out using a SQUID magnetometer in the temperature range 2 K to 300 K at applied field of 10 kOe. Conductivity measurements were performed on pellets using a standard four-probe measurement setup. Potassium cobalt hydrate was stored in humidified containers to avoid dehydration and decomposition.

## 3. Results and Discussion

During the oxidation and water intercalation of $K_{0.6}CoO_2$ using $K_2S_2O_8$ in an aqueous solution significant Bragg peak shifts of the *00l* reflections towards larger d-spacings are observed, while preserving the hexagonal symmetry ($P6_3/mmc$). The stoichiometry of the MLH $K_{0.3}CoO_2 \cdot 0.4H_2O$ was obtained by ICP-AES and TGA. Fig. 2 shows TGA data of $K_{0.3}CoO_2 \cdot 0.4H_2O$ on heating at 0.25°C/min in flowing Ar (g). The completely dehydrated $K_{0.3}CoO_2$ phase was obtained after a weight loss of about 5 %. This corresponds to ~ 0.4 water per formula unit and is consistent with a single potassium-water layer separating the $CoO_2$-layers.

The capillary was spun during the powder diffraction measurement to ensure good powder averaging. The first two peaks in the resulting powder pattern are indicative of the interlayer spacing similar to what was observed in the monolayer hydrate $Na_xMO_2 \cdot yH_2O$ (M = Co, Rh, Ru) [5,6,7]. The MLH structures in Fig. 1 where $H_2O$ and cation sites are partially occupied within a single layer between the $MO_2$ layers are shown. Subsequent indexing of the $K_{0.3}CoO_2 \cdot 0.4H_2O$ powder pattern was consistent with a hexagonal cell with axis lengths ($a$ = 2.8262(1), $c$ = 13.8269(6) Å), which are slightly smaller than those ($a$ = 2.8344(7), $c$ = 13.842(5) Å) of the hexagonal $Na_{0.36}CoO_2 \cdot 0.7H_2O$ analogue. The interlayer spacing of $K_{0.3}CoO_2 \cdot 0.4H_2O$ (6.91 Å) is also slightly reduced compared to the one observed for $Na_{0.36}CoO_2 \cdot 0.7H_2O$ (6.92 Å). The lower hydration level and different Co oxidation state (see below) in $K_{0.3}CoO_2 \cdot 0.4H_2O$ result in smaller lattice constants as well as a reduced interlayer spacing in comparison to $Na_{0.36}CoO_2 \cdot 0.7H_2O$. The structure was refined using the Rietveld method [13,14]. Asymmetry corrections using the Finger, Cox and Jephcoat's formalism [15,16] were then introduced to the model diffraction peak shapes, and a March-Dollase ellipsoid model was used to account preferred orientations [17].. The occupancies for the potassium cations and water molecules, which occupy the same site, were fixed based on the ICP and TGA results. An ordered K/H$_2$O model as the one in Na-MLH leads to a poorer fit with the data (Rwp= 9.8% vs 8.2%). K cannot be located in the position 6h (1/3,2/3,1/4) as this results into too short K-O distances within the gallery height. A shift off these special positions is thus required. The final profile fit based on the disordered model listed in table 1 depicted in Fig. 3. The inset shows a comparison of the Na-MLH and K-MLH structures. The refined structural model and selected interatomic distances are also summarized in Table 1. The structure consists of layers of edge-sharing cobalt oxide octahedra with disordered water molecules and potassium cations located on the same site intercalated in-between them. The Co-O distance (1.868(1) Å) found in the octahedral layers is slightly shorter than the one observed in Na$_{0.36}$CoO$_2$·0.7H$_2$O (1.878 Å). This is consistent with a higher oxidation state of the Co [18] in K$_{0.3}$CoO$_2$·0.4H$_2$O and also reflected ,as mentioned above, in the smaller unit cell parameters. The extraction of K$^+$ ions using K$_2$S$_2$O$_8$ thus leads to a slightly more oxidized CoO$_2$ layer.

A SQUID magnetometer (Quantum Design) was used to determine the magnetic susceptibility (M/H) of the potassium oxide hydrate as a function of temperature. Nakamura and co-workers have reported that a peak in the temperature dependence of the dc magnetic susceptibility of K$_{0.5}$CoO$_2$ is located around 35 K corresponding to the Néel temperature of Co$_3$O$_4$ (T$_N$ = 33K), which is present due to unreacted Co$_3$O$_4$ in the sample [9]. We have found this peak in the susceptibility data for both K$_{0.6}$CoO$_2$ and K$_{0.3}$CoO$_2$·0.4H$_2$O (Fig. 4) despite the fact that the X-ray diffraction pattern of K$_{0.6}$CoO$_2$ and K$_{0.3}$CoO$_2$·0.4H$_2$O do not show any indications of crystalline Co$_3$O$_4$. We argue that both peaks observed in the magnetization are due to Co$_3$O$_4$ impurities and not of intrinsic nature. K-MLH shows paramagnetic behavior similar to K$_{0.5}$CoO$_2$ [9]. The temperature dependence of the resistivity for pressed disks of K$_{0.3}$CoO$_2$·0.4H$_2$O and Na$_{0.3}$CoO$_2$·0.7H$_2$O is shown in Fig. 5. K$_{0.3}$CoO$_2$·0.4H$_2$O shows a metallic behavior as the resistivity, of the order 10 m·Ω·cm, is linearly decreasing with decreasing temperature as shown in Na$_{0.3}$CoO$_2$·0.7H$_2$O.

**Acknowledgement**

This work was supported by an LDRD from BNL. Research carried out in part at the NSLS at BNL is supported by the U.S. DOE (DE-Ac02-98CH10886 for beam line X7A).

**Figure Captions**

**Fig. 1:**
Crystal structures of (a) $Na_{0.36}CoO_2 \cdot 0.7H_2O$, (b) $Na_{0.3}RhO_2 \cdot 0.6H_2O$, and (c) $Na_{0.22}RuO_2 \cdot 0.45H_2O$. Lines in the figures define the hexagonal unit cells.

**Fig. 2:**
Thermogravimetric analysis (TGA) of $K_{0.3}CoO_2 \cdot yH_2O$ determined with a heating rate $0.25^{o}C/min$ under Ar (g) flowing.

**Fig. 3:**
Results of Rietveld refinements of the structural model of the $K_{0.3}CoO_2 \cdot 0.4H_2O$ at room temperature using synchrotron X-ray powder diffraction data. Marks below the data indicate the positions of the symmetry allowed reflections. The lower curve represents the difference between observed and calculated profiles ($I_{obs} - I_{calc}$) plotted on the same scale as the observed data. (Inset) The structures of (a) $K_{0.3}CoO_2 \cdot 0.4H_2O$ and (b) $Na_{0.36}CoO_2 \cdot 0.7H_2O$ viewed perpendicular to the c-axis are shown. Lines in the figures define hexagonal unit cell of $K_xCoO_2$ and $K_{0.3}CoO_2 \cdot 0.4H_2O$, respectively.

**Fig. 4:**
Temperature dependence of the dc magnetic susceptibility of $K_{0.6}CoO_2$ (circles) and $K_{0.3}CoO_2 \cdot 0.4H_2O$ (triangles).

**Fig. 5:**
The resistivity as a function of temperature for $K_{0.3}CoO_2 \cdot 0.4H_2O$ and $Na_{0.3}CoO_2 \cdot 0.7H_2O$.

**Figure 1**

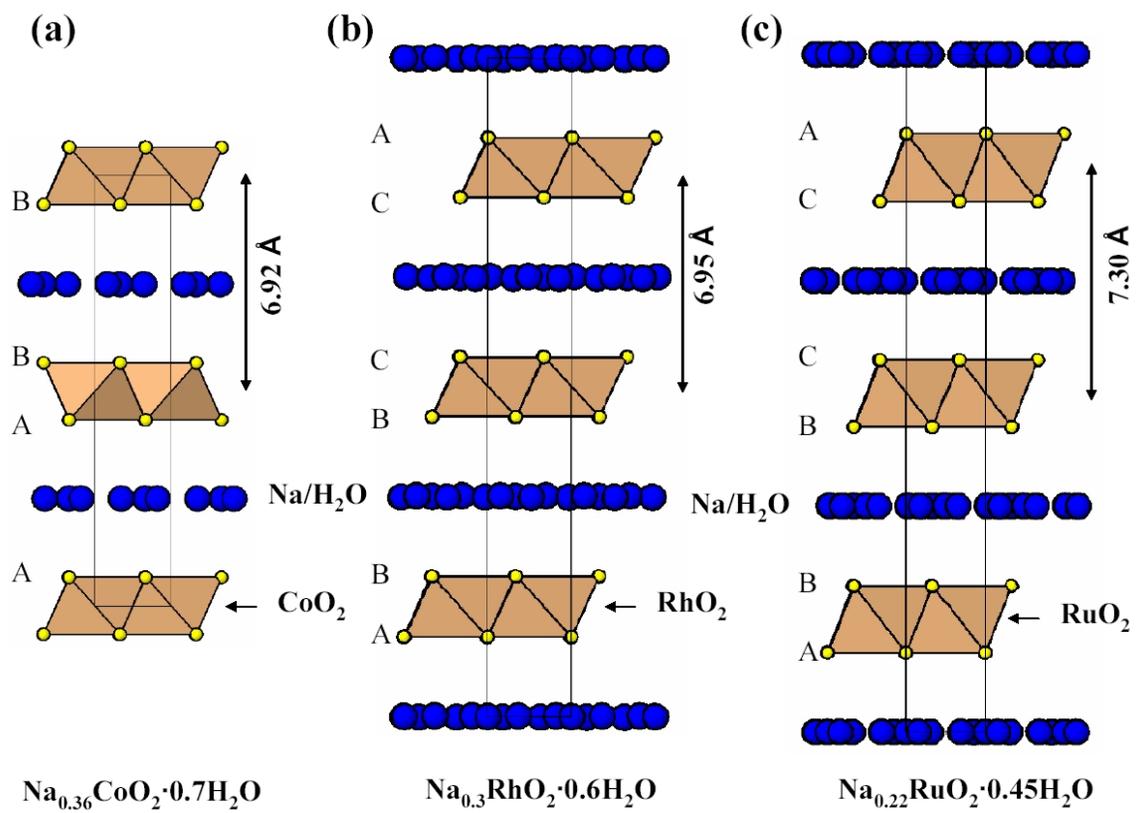

**Figure 2**

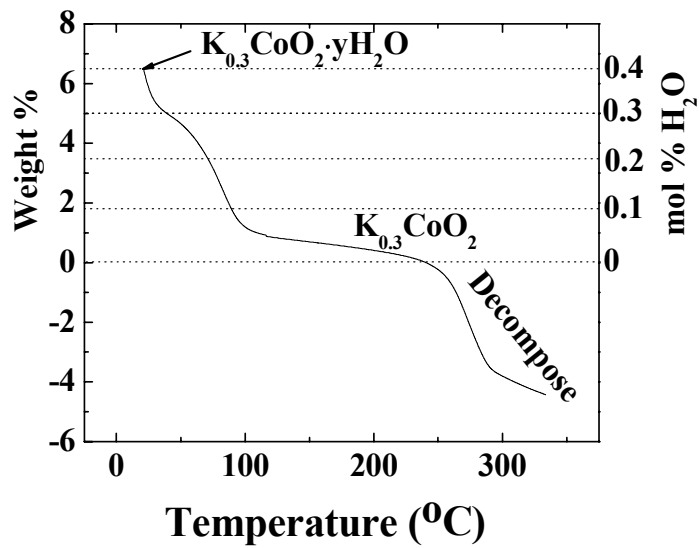

**Figure 3**

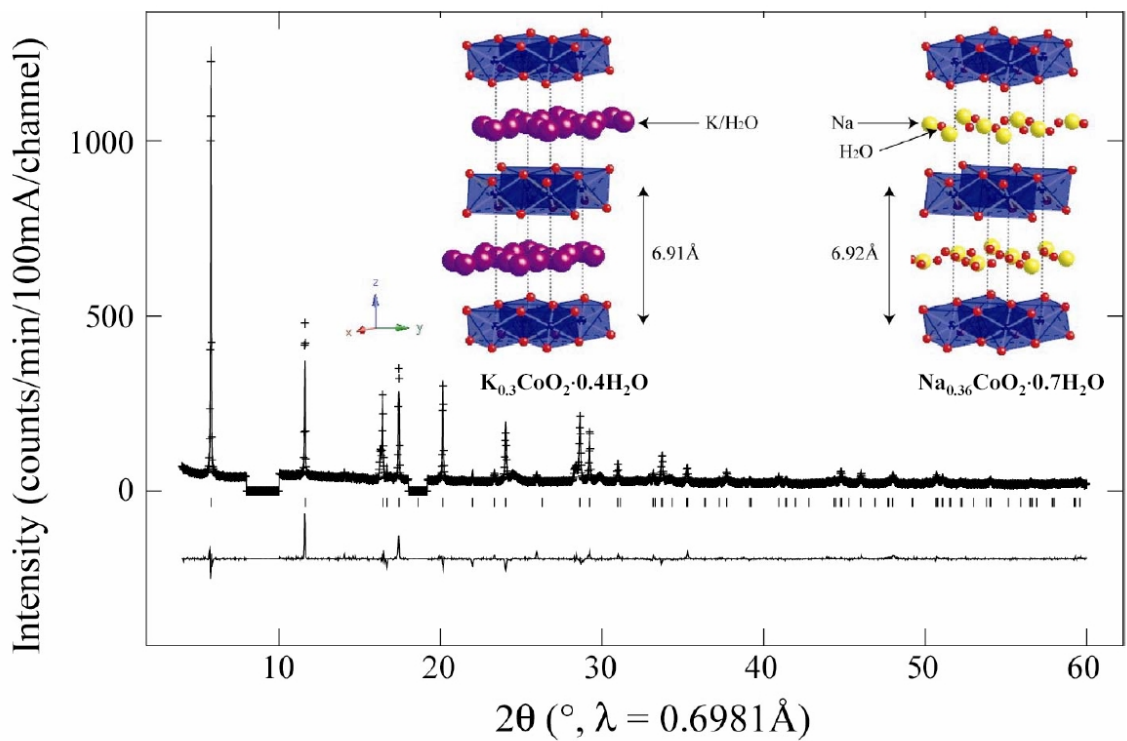

**Figure 4**

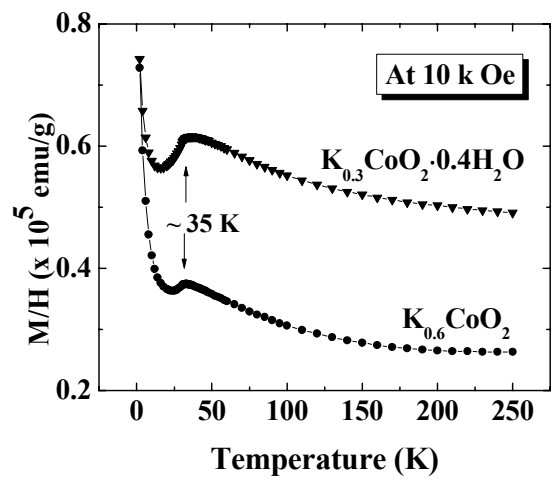

**Figure 5**

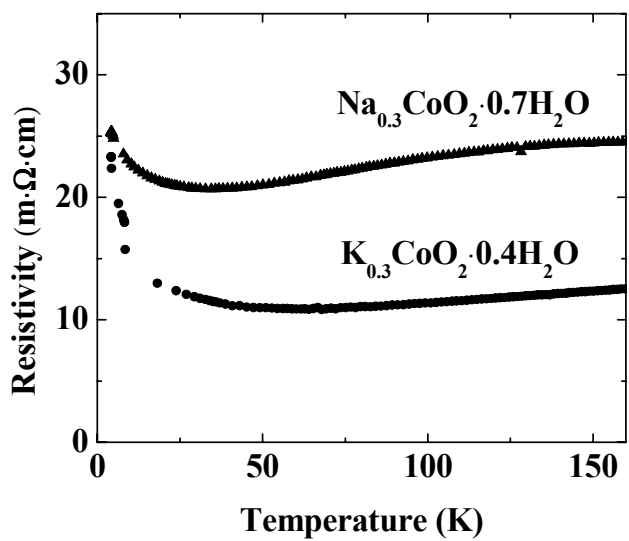

Table 1. Atomic coordinates of $K_{0.3}CoO_2 \cdot 0.4H_2O$.

| Atom | Wyckoff | x | y | z | occu. | $U_{iso}$ (Å$^2$) |
|---|---|---|---|---|---|---|
| Co | 2a | 0 | 0 | 0 | 1 | 0.0083(4) |
| O | 4f | 1/3 | 2/3 | 0.0658(2) | 1 | 0.057(3) |
| K | 6h | 0.164(2) | 2x(K) | 1/4 | 0.1 | 0.044(6) |
| OW | 6h | x(K) | 2x(K) | 1/4 | 0.133 | $U_{iso}$(K) |

Space group $\boldsymbol{P6_3/mmc}$; $a$ = 2.8262(1) Å, $c$ = 13.8269(6) Å, $V$ = 95.645(6) Å$^3$. $\chi^2$ = 6.7, $_wR_p$ = 8.24%, $R_p$ = 5.83%, $R_{F2}$ = 15%., $R_e$=6.6%  OW denotes water molecule modeled with oxygen atom. Restraints were used to tie isotropic atomic displacement parameters same for the interlayer species. Selected interatomic distances (Å) are Co–O = 1.868(1)Å (×6), K-O = 2.678(4)Å (×2), 3.333(4)Å (×4).